\documentclass[12pt]{article}
\newcommand{\ba}{\begin{eqnarray}}
\newcommand{\ea}{\end{eqnarray}}

\newcommand{\id}
 {i\kern.06em\hbox{\raise.25ex\hbox{$/$}\kern-.60em$\partial$}}
\newcommand{\as}{/\kern-.52em A}
\newcommand{\Ds}{/\kern-.69em D}
\title{On the two-dimensional Fermion Determinant
 at Finite Temperature}
\author
{C.D.~Fosco$^a$\thanks{CONICET, Argentina}\,,
R.E.~Gamboa Sarav\'\i$\,^{b\, *}$\,and\,
 F.A.~Schaposnik$^b$\thanks{Investigador CICBA}
\\
{\normalsize\it $^a$Centro At\'omico Bariloche, 8400 Bariloche,
Argentina}\\
{\normalsize\it $^b$Departamento de F\'\i sica, Universidad
Nacional de La
Plata}\\ {\normalsize\it C.C. 67, 1900 La Plata, Argentina}
}

\begin{document}
\maketitle
\date{}
\begin{abstract}
We evaluate the fermionic determinant for massless $QED_2$ at finite 
temperature, in the imaginary time formalism. By using a decoupling 
transformation of the fermionic fields, we show that the determinant 
factorizes into the usual, temperature independent expression, times an 
extra factor which depends on the temperature and on the constant  
component of the gauge field. 
\end{abstract}
\bigskip
\newpage

Recent results on the finite temperature effective action for
fermions in a gauge field background $A_\mu$ in  $0+1$ \cite{DLL}-\cite{BD},
$2+1$ \cite{DGS}-\cite{PS}, $1+1$ \cite{DS} and $d+1$ \cite{FRS3}
dimensions have revealed new and interesting
features, showing in particular
 that $S_{eff}[A_\mu] $ can be a non-extensive quantity
depending on $A_\mu$ in a non-polynomial way.

In the present note we reconsider the $1+1$ case, that is, we analyse
the effective action for the massless Schwinger Model at finite temperature,
following a covariant approach based in the introduction of a heat-bath
velocity $u_\mu$ \cite{W}. In fact, the fermion determinant for two dimensional
 massless fermions in a gauge field background  has been computed 
 using different techniques both for topologically trivial and non trivial
backgrounds (where zero modes have to be carefully handled)
\cite{L}-\cite{DW}. Here, we want to clarify, in connection with
 the results of
 \cite{DS},  the way in which the heat bath velocity 
can be connected with  non-trivial boundary conditions leading to a result 
which is related 
to that obtained in \cite{SW} in the analysis of the
fermion determinant considering
zero modes in the various
topological sectors. As we shall see, taking into account 
the heat-bath velocity  is tantamount to considering a constant
gauge field background (or, equivalently, twisted boundary conditions for
the fermions and no constant background).
We will find that there is a contribution to the fermion
determinant which depends on $u_\mu$, this showing that in general 
the Schwinger Model effective action is not a trivial extension of the
zero temperature one.

~

We shall begin by dealing with ${\cal Z}[A]$, the two dimensional  
fermionic
determinant in the presence of an Abelian $U(1)$  gauge field
in the imaginary time formalism

\begin{equation}\label{fdet}
  {\cal Z} [A] \;=\; \det [\id + i e \not \!\!A ]
  \;=\;\int {\cal D}{\bar \psi} {\cal D}\psi\,
  \exp \left[ - S_F({\bar \psi},\psi;A) \right]
\end{equation}
where the action functional $S_F$ is given by
\begin{equation}\label{fact}
  S_F[{\bar \psi},\psi;A] \;=\; \int_0^\beta d\tau \int d x \,
 {\bar \psi}(\tau,x) [\not \! \partial + i e \not \!\! A(\tau,x) ]
 \psi (\tau,x)
\end{equation}
where $\tau$ denotes the imaginary time, $x$ the (sole) space  
coordinate,
and $\beta = \frac{1}{T}$. Spacetime is Euclidean, and the fields  
obey the standard boundary conditions at $T>0$
 $$A_\mu (\beta,x) \;=\; A_\mu (0,x) \;\;\; \forall x  
\;\;\mu\,=\,1,2 \;,$$
\begin{equation}\label{bcn}
  \psi (\beta,x) \;=\; - \psi (0,x) \;\;,\;\;
  {\bar\psi} (\beta,x) \;=\; - {\bar\psi} (0,x)\;\;\; \forall x \;.
\end{equation}

It is a well established result \cite{L}-\cite{DW} 
that at finite temperature this 
fermionic determinant is formally identical to the zero temperature one, 
but only if the (often non sufficiently emphasized) assumption is  
made that the external gauge field is topologically trivial in the sense 
that its holonomy around the periodic time coordinate is trivial. We shall  
show that relaxing this assumption leads to a more general answer. 
Moreover, it should of course lead to the usual result when
the same kind of configuration is considered.

We shall decompose the external gauge field $A_\mu$ as follows
\begin{equation}\label{dsca}
  A_\mu \;=\; \frac{1}{e} ( \partial_\mu \varphi + i \epsilon_{\mu\nu} 
  \partial_\nu \sigma ) \;.
\end{equation}

It follows from eq.(\ref{dsca})  that the scalar fields $\varphi$  
and $\sigma$
satisfy the equations
\begin{equation}\label{dsc1}
  \partial^2 \varphi \;=\; e \,\partial \cdot A \;\;\;
  \partial^2 \sigma \;=\; i e \,\epsilon_{\mu\nu} \partial_\mu A_\nu \;.
\end{equation}

It must be noted that equations (\ref{dsc1}) only determine the fields 
$\varphi$ and $\sigma$ up to a solution of Laplace equation. Namely,

\begin{eqnarray}\label{gsol}
  \varphi (\tau , x) &=& \varphi^{(0)}(\tau,x) + {\tilde
  \varphi}(\tau,x)\nonumber\\
  \sigma  (\tau , x) &=& \sigma^{(0)}(\tau,x) + {\tilde \sigma}(\tau,x) 
\end{eqnarray}
with
\begin{equation}\label{lpl}
 \partial^2 \varphi^{(0)}(\tau,x) \;=\;0 \;\;,\;\; \partial^2  
\sigma^{(0)}(\tau,x)\;=\;0
\end{equation}
and ${\tilde \varphi}$, ${\tilde \sigma}$ particular solutions of
(\ref{dsc1}).

The new, non trivial, part of $A_\mu$,
playing  a central role in our calculation,
 may be thought of as coming from  
$\varphi^{(0)}$ and $\sigma^{(0)}$: indeed, at $T=0$ there is no  
reason to
inlude non-trivial solutions of the Laplace equations, since the  
spacetime
manifold is trivial, and a regular solution is necessarily a constant 
(giving no contribution to $A$). Of course, Poincar\'e invariance also 
forbids the imposition of any non trivial boundary  condition at  
infinity.
On the other hand, at finite temperature the topology of spacetime  
is the
one of a cylinder, and moreover there is no Poincar\'e invariance. The 
partition function of a system, in the canonical ensemble say, is  
defined
in terms of the Hamiltonian, a non-covariant object. Moreover, the
definition of this ensemble implicitly assumes the existence of a
preferential reference frame, the one where the thermal bath is at rest 
with respect to the system. When there is relative motion between the 
system and the bath, we should expect the results to depend not  
only on the energy, but also on the other available integrals of motion 
of the system: for example its momentum~\cite{lan}, which is simply 
related to the velocity~\footnote{As the total fermionic charge is conserved, 
one may also introduce the total current.}. 

This implies that we may now write the general solutions to (\ref{lpl}) 
using a constant vector, the velocity $u_\mu$ with respect to the  
bath. It is not hard to realize that the most general solution (compatible  
with the periodicity of $A$ in the time direction) is a linear function of the 
coordinates. The only available vector coefficient to build up a  
scalar is something proportional to $u_\mu$:
\begin{equation}
\varphi^{(0)}\;=\; a \; u_\mu x_\mu  \;\;,\;\; \sigma^{(0)} \;=\; b \; 
u_\mu x_\mu
\end{equation}
where $a$ and $b$ are constants. The relation between the  
coefficients $a$
and $b$ and $A_\mu$ is determined by the equations
\begin{equation}\label{relt}
  A_\mu^{(0)} \;=\; \frac{1}{e}
  ( a \, u_\mu \;+\; i\, b \,\epsilon_{\mu\nu} u_\nu ) \;\;
  \mu = 0 , 1 \;,
\end{equation}
where $A_\mu^{(0)}$ denotes the zero momentum (i.e., constant)  
component of
the gauge field.

With this remark in mind, we see the gauge field $A$ appearing in the 
fermionic determinant is of the form
\begin{equation}
A_\mu \;=\; A_\mu^{(0)} + {\tilde A}_\mu
\end{equation}
where
\begin{equation}
{\tilde A}_\mu \;=\; \frac{1}{e} ( \partial_\mu {\tilde \varphi} + i 
\epsilon_{\mu \nu} \partial_\nu {\tilde \sigma}) \;.
\end{equation}
Of course, ${\tilde A}$ has no zero momentum component, and in  
consequence
the scalar fields ${\tilde \varphi}$ and ${\tilde \sigma}$ are strictly 
periodic in  the time coordinate $\tau$. It should be noted that the "new" 
part of the gauge field we are including in this study is, in Fourier space,  
proportional to a delta function of Euclidean momentum. This suggests a  
connection with the perturbative study by Das et al in real time, where the 
extra piece found in the effective action has support for $k^2 = 0$. This 
support, when mapped to Euclidean spacetime, becomes $k_\mu=0$.

The ${\tilde A}$ part of the gauge field may be entirely decoupled  
by the anomalous Jacobian method, just by noting that
\begin{equation}
S_F ({\bar \psi},\psi;A) \;=\; \int_0^\beta d \tau \int dx \, {\bar 
\psi}(\tau,x) e^{-i e ({\tilde \varphi} - \gamma_5 {\tilde  
\sigma})} ( \not
\! \partial + i e \not \! A^{(0)} )
 e^{i e ({\tilde \varphi} + \gamma_5 {\tilde \sigma})}
\psi (\tau,x)
\end{equation}
and defining the new fermionic fields
\begin{eqnarray}
\chi \, & = &\,  \exp \left( {i e ({\tilde \varphi} + \gamma_5  
{\tilde \sigma})}
\right)\psi
\nonumber\\ {\bar \chi}  \,& = & \, {\bar \psi} \exp \left( {-i e  
({\tilde \varphi} -
\gamma_5 {\tilde \sigma})} \right) \label{tra}
\end{eqnarray}
Then, one can write
\begin{equation} \label{Z}
{\cal Z}(A) \;=\;   J(A^{(0)},{\tilde A})   \times  {\cal Z} (A^{(0)}) 
\end{equation}
where $J(A^{(0)},{\tilde A})$ denotes the anomalous Jacobian for the 
transformation (\ref{tra}).

Eq.(\ref{Z})  represents one of the main steps in our calculation:  
we have
managed to factorize the fermion determinant into one factor which
corresponds to the usual Jacobian leading, as we shall see, to
the usual Schwinger determinant  but in a domain $(0,L) \times(0,\beta)$,
times a determinant in a constant background which is in fact related to
the velocity $u_\mu$ with respect to the bath.
(For the sake of generality we also assume that the space like coordinate
is finite, with length $L$, and the fermions are periodic in this 
direction.This shall also be important when considering the thermodynamic  
limit).

Now we shall  show that $J$ is actually {\em independent of
$A^{(0)}$ \/}, what makes it identical to the standard, zero temperature 
like result (with the imaginary time integral in the
domain $(0,\beta)$). To show this property we only have to realize  
that the
Jacobian for an infinitesimal axial transformation is independent of 
$A^{(0)}$, since the finite transformation is built as an iteration of 
infinitesimal ones. The infinitesimal axial transformation
\begin{equation}
\psi \to \psi + i \alpha \gamma_5 \psi \;\;,\;\; {\bar \psi} \to {\bar 
\psi} + i \alpha {\bar \psi} \gamma_5 \;,
\end{equation}
induces a Jacobian
\begin{equation}
J \;=\; \exp [i {\cal A} ]
\label{jj}
\end{equation}
with
\begin{equation}
{\cal A} \;=\; {\rm Tr} [ \alpha \gamma_5 ]
\label{jj1}
\end{equation}
where the trace is meant to be on Dirac indices, as well as on  
functional
space. As usual, to make sense of ${\cal A}$, we define it as the  
limit of
a regularized expression:
\begin{equation}\label{arg}
{\cal A} \;=\; \lim_{M \to \infty} {\rm Tr} [ \alpha \gamma_5 f
(\frac{{\Ds}^2}{M^2}) ]
\label{jj2}
\end{equation}
where $f$ is a function that verifies
\begin{equation}
f(0) \,=\,1 \;\;,\;\; f(\infty) \,=\, f'(\infty) \,=\, \cdots = 0 \;. 
\label{jj3}
\end{equation}
Now we write the trace in terms of eigenstates of the free Dirac  
operator,
which are free spinors, with a discrete time component for the momentum 
\begin{equation}\label{arg1}
{\cal A} \;=\; \lim_{M \to \infty} {\rm tr} \left\{ \frac{1}{\beta}  
\sum_n
\int \frac{d k}{2 \pi} \, \langle n , k | [\alpha \gamma_5 f(\frac{{
\Ds}^2}{M^2}) ] |n , k \rangle \ \right\} \;.
\end{equation}
The sum over $n$ can be put as an integral over a
continuum momentum, plus a temperature dependent term which picks up 
contributions from the single poles of $f$. We shall assume that  
$f$ is a
meromorphic function with no singularities on the real axis. The
temperature dependent term obviously vanishes in the limit $M \to  
\infty$,
since there is one integration over momentum less than required to  
give a
non-zero contribution. Indeed, this is what happens in the usual  
proof of
the temperature independence of the anomaly. Thus we are lead to an 
expression like (\ref{arg1}) but with a double integral rather that  
a sum
and an integral. Yet it is not indentical to the zero temperature  
Jacobian,
since ${\Ds}^2$ depends on the constant component of $A$,
\begin{equation}
{\Ds}^2 \,=\, D^2(A) + \frac{i e}{4} [\gamma_\mu,\gamma_\nu]
F_{\mu\nu} ({\tilde A})
\label{jj4}
\end{equation}
where the only dependence on $A^{(0)}$ comes from $D^2(A)$. But this 
dependence is erased by a simple shift of the (now continuous) timelike 
momentum, thus the Jacobian is independent of $A^{(0)}$.

After some standard calculation, we then have from  
eqs.(\ref{jj})-(\ref{jj4})
\begin{eqnarray}
 J[A^{(0)},\tilde A] = \exp \left( -\frac{e^2}{2\pi} \int_o^\beta  
d\tau \int_0^L dx
\tilde A_\mu \Delta^{\mu \nu} \tilde A_\nu \right)
\label{J1}
\end{eqnarray}
with $\Delta_{\mu \nu}$ given by
\begin{eqnarray}
\Delta_{\mu \nu} = \delta_{\mu \nu} - \partial_\mu \partial^{-\!2}  
\delta_\nu
\label{J2}
\end{eqnarray}

There only remains
to evaluate the constant field determinant
\begin{equation}
{\cal Z} (A^{(0)}) \;=\; \det [ \not \! \partial + i e \not \!  
A^{(0)}] \;.
\end{equation}

It is immediate to realize that, as the fermions are massless, the
determinant can be factorized into two constant field determinants,  
one for
each chirality
\begin{equation}
{\cal Z} ( A^{(0)}) \;=\; \det (D_0 + i D_1) \det (D_0 - i D_1)  
\exp(-q) \;,
\end{equation}
where $q$ is a counterterm  which can be identified with the non  
holomorphic residue
\cite{AMV}. Each one of the chiral determinants can be
 exactly calculated following for
example the technique described in this last reference so that
one  arrives to the result
\begin{equation}
{\cal Z} (A^{(0)}) \;=\; \exp [- \Gamma ( A^{(0)})]  \;,
\end{equation}
with
\begin{equation}\label{gdef}
\Gamma ( A^{(0)}) \,=\, - \log \left|\frac{\vartheta (\alpha , \tau)}{ 
\vartheta (0,\tau)}\right|^2 \,+\,  \frac{e^2}{2\pi}\beta L
{ A}^{(0)}_\mu { A}^{(0) \mu}
\end{equation}
where the Jacobi Theta function $\vartheta$  may be defined by a  
single series
representation
\begin{equation}
\vartheta (\alpha , \tau) \;=\; \sum_n \, e^{ i \pi \tau n^2 \,+\,  
2 \pi i
n \alpha} \;, \label{theta}
\end{equation}
and
\begin{equation}
\tau \;=\; i \frac{L}{\beta} \;\;,\;\; \alpha \;=\;
-\frac{L}{2\pi}\left({A}^{(0)}_1 - i { A}^{(0)}_0 \right)\,.
\end{equation}

The first term in the r.h.s. of eq.(\ref{gdef}) corresponds to the  
product of holomorphic and antiholomorphic
factors arising in the computation of chiral determinants in a  
constant background.
The second term is just $q$ and
represents the obstruction to holomorphic factorization.
It can be computed just by demanding  the complete determinant  
to be  gauge-invariant.

We then have, putting together (\ref{J1}) and (\ref{gdef}) that  
${\cal Z}(A)$ as given by
(\ref{Z}) can be written in the form
\begin{eqnarray}
{\cal Z}(A) &= & \exp \left( -\frac{e^2}{2\pi} \int_o^\beta d\tau  
\int_0^L dx
\tilde A_\mu \Delta^{\mu \nu} \tilde A_\nu \right) \exp\left(  
-\frac{e^2}{2\pi}\beta L
{ A}^{(0)}_\mu { A}^{(0) \mu}
\right) \times
\nonumber\\
& & \left|\frac{\vartheta (\alpha , \tau)}{
\vartheta (0,\tau)}\right|^2
\label{una}
\end{eqnarray}
or, in view of the boundary conditions, more compactly as
\begin{equation}
{\cal Z}(A) =  \exp
\left(
-\frac{e^2}{2\pi} \int_o^\beta d\tau \int_0^L dx
A_\mu \Delta^{\mu \nu}  A_\nu
\right)
\left |
\frac{\vartheta (\alpha , \tau)}
{\vartheta (0,\tau)}
 \right |^2
\label{red}
\end{equation}
which is our main result. It is worth emphasizing a nontrivial property
satisfied by this expression, namely, that the contributions coming from
topological (non-zero constant field component) and non-topological gauge
field configurations are neatly decoupled. This is in fact a consequence
of the independence of the chiral anomaly on the long distance properties
of the system.   
This decoupling is in fact also observed in the real time formulation
calculation presented in (\cite{DS}), where one sees that the difference
between the result presented therein and the usual is non-vanishing only
for gauge fields having support on the light cone. 

We note that to find the {\em explicit\/} form of relation (\ref{relt}) 
would require the calculation, for example, of the expectation value of 
the total momentum as a function of $A$.

\underline{Acknowledgments}:
F.A.S. thanks  the Laboratoire 
de Physique Th\'eo\-ri\-que et Hautes Energies de l'Universit\'e
de Paris VII for kind hospitality during part of this
work.  C.D.F. thanks the members of the Department of Physics of the 
University of Oxford where part of his work on this subject was done.
We acknowledge A.J. da Silva for some useful comments.
This work is
partially  supported by CICBA, CONICET (PIP 4330/96), ANPCyT grants
(PICT 97-2285) and 03-00000-02249, and
Fundaci\'on Antorchas,   Argentina.
\newpage


\end{document}